# Parity transitions in the superconducting ground state of hybrid InSb-Al Coulomb islands

Jie Shen[1]*[†], Sebastian Heedt[1]*, Francesco Borsoi[1]*, Bernard van Heck[2], Sasa Gazibegovic[1,3], Roy L. M. Op het Veld[1,3], Diana Car[1,3], John A. Logan[4], Mihir Pendharkar[5], Senja J. J. Ramakers[1], Guanzhong Wang[1], Di Xu[1], Daniël Bouman[1], Attila Geresdi[1], Chris J. Palmstrøm[4,5], Erik P. A. M. Bakkers[1,3] & Leo P. Kouwenhoven[1,6†]

[1]QuTech and Kavli Institute of Nanoscience, Delft University of Technology, 2600 GA Delft, The Netherlands.

[2]Microsoft Quantum, Microsoft Station Q, University of California Santa Barbara, Santa Barbara, California 93106-6105, USA

[3]Department of Applied Physics, Eindhoven University of Technology, 5600 MB Eindhoven, The Netherlands.

[4]Materials Department, University of California Santa Barbara, Santa Barbara, California 93106, USA.

[5]Electrical and Computer Engineering, University of California Santa Barbara, Santa Barbara, California 93106, USA.

[6]Microsoft Station Q at Delft University of Technology, 2600 GA Delft, The Netherlands.

*These authors contributed equally to this work.

†J.Shen-1@tudelft.nl, Leo.Kouwenhoven@Microsoft.com

**Abstract**

**The number of electrons in small metallic or semiconducting islands is quantized. When tunnelling is enabled via opaque barriers this number can change by an integer. In superconductors the addition is in units of two electron charges (2$e$), reflecting that the Cooper pair condensate must have an even parity. This ground state (GS) is foundational for all superconducting qubit devices. Here, we study a hybrid superconducting–semiconducting island and find three typical GS evolutions in a parallel magnetic field: a robust 2$e$-periodic even-parity GS, a transition to a 2$e$-periodic odd-parity GS, and a transition from a 2$e$- to a 1$e$-periodic GS. The 2$e$-periodic odd-parity GS persistent in gate-voltage occurs when a spin-resolved subgap state crosses zero energy. For our 1$e$-periodic GSs we explicitly show the origin being a single zero-energy state gapped from the continuum, i.e. compatible with an Andreev bound states stabilized at zero energy or the presence of Majorana zero modes.**

**Introduction**

A superconductor can proximitize a semiconductor and open a gap in its energy spectrum. If the two materials are strongly coupled, the induced gap can be as large as the original gap in the superconductor. The two gaps respond differently to an applied magnetic field, e.g. when the Landé $g$-factors differ in the two materials. A large $g$-factor in the semiconductor can cause the induced gap to close long before the closing of the original gap. If, in addition, the semiconductor has strong spin-orbit interaction, the induced gap can re-open, signalling a transition to a topological superconducting phase [1,2]. This phase contains pairs of Majorana zero modes (MZMs) that can accommodate either zero or one fermion, and thus allows for both even- and odd-parity GSs [3,4].

When a conductor has a finite size, it forms an island restricting the charge to an integer times the elementary charge, $e$ [5]. The resulting Coulomb blockade effects have been widely studied in metallic and superconducting islands, the latter often referred to as Cooper pair boxes [6,7]. A major breakthrough was the demonstration of charge quantization in units of $2e$ in aluminium (Al) islands [8-13], indicating that the even-parity superconducting GS was not poisoned by quasiparticles on the time scale of the measurement. The $2e$ quantization could be destroyed by subjecting the Al to an external magnetic field, $B$, which causes a transition to the metallic state with $1e$ charge quantization [10,11].

Hybrid superconducting-semiconducting islands have also shown a $2e$ charge quantization at low $B$-fields [14-16]. These observations imply that the low-energy spectrum in the semiconductor is completely proximitized with no Andreev bound states (ABSs) at low energies. Also, for these hybrid islands a $B$-field can cause a $2e$ to $1e$ transition [15,16]. A recent breakthrough demonstrated that under particular circumstances the $1e$ quantization is not due to the transition to the metallic state - but rather due to a topological superconducting phase [16]. These pioneering experiments used InAs as the semiconductor.

Here, we harness the large $g$-factor ($g \approx 50$) and the ballistic transport properties of InSb nanowires [17,18], and find additional $B$-field induced transitions, including a recurrence of a $2e$ quantization at higher $B$-fields.

**Results**

**Different *B*-field GS evolutions at controllable gate configurations**

Our device (Fig. 1a) consists of a hexagonal InSb nanowire with two of its facets covered by a thin epitaxial layer of Al (see ref. [19] for materials details). Two top gates (TG) can induce adjustable tunnel barriers separating the InSb-Al island from the two normal leads. The voltage, $V_{PG}$, applied to the top plunger gate (PG), can be used to tune the charge on the island as well as the spatial charge density profile in the semiconductor. A bias voltage, $V_b$, is applied between source (S) and drain (D), yielding a current, *I*, that is measured in a dilution refrigerator at a base temperature of ~15mK.

Fig. 1b shows the differential conductance, $dI/dV_b$, versus $V_b$ and the voltage applied to the tunnel gate at *B* = 0. The charge is fixed in the current-blockaded Coulomb diamonds (with mostly blue colour) with a periodicity in gate voltage corresponding to a charge increment of 2*e*. For $V_b$ > 120 µV the periodicity is halved to 1*e*, indicating the onset of single electron transport. Linecuts in the right panel show that $dI/dV_b$ can be enhanced as well as suppressed, even down to negative values (black colour). These are known features for hybrid islands and can be used to extract values for the charging energy, $E_c = e^2/2C \approx 25$ µeV and the lowest-energy subgap state, $E_0$ = 50-90 µeV (Supplementary Fig. 2). The Al superconducting gap, $\Delta$ = 220 µeV, is extracted from tunnelling spectroscopy measurement (Supplementary Fig. 3).

To understand parity transitions induced by a *B*-field, we illustrate four different scenarios in Figs. 1c-f. We sketch a slow reduction of $\Delta$ for a *B*-field, $B_\parallel$, along the nanowire axis (Figs. 1c-e) and a rapid decrease of $\Delta$ for a perpendicular field, $B_\perp$ (Fig. 1f). We consider a single lowest-energy subgap state (i.e. an ABS) with energy $E_0$, which is two-fold spin-degenerate at $B = 0$ and becomes spin-split in a $B$-field. Fig. 1c sketches the case where $E_0$ decreases very slowly with $B_\parallel$, remaining above $E_c$ such that the GS parity remains even. This translates to $2e$-periodic conductance oscillations for all $B_\parallel$-fields (Fig. 1k). Note that conductance peaks occur when the lowest-energy parabolas cross. In Fig. 1g the lowest crossings are always between even-charge parabolas. These crossings, for instance at $N_g$ = -1 and +1, are $2e$-periodic. The odd-charge parabolas remain above these lowest-energy crossings and thus do not participate in the low-energy transport.

Fig. 1d sketches a second case where $E_0$ varies more rapidly with $B_\parallel$. When $E_0$ crosses $E_c$ the odd-charge parabolas pass the lowest-energy crossings of the even parabolas, thereby adding degeneracy points between even-charge and odd-charge parabolas (Fig. 1h). This results in alternating smaller and larger peak spacings, where the smaller valleys have odd-parity for $E_0 > 0$ and even-parity for $E_0 < 0$. Note that an equal spacing with $1e$-periodicity occurs when $E_0 = 0$. At negative energies, when $E_0$ crosses -$E_c$ the $2e$-periodicity is restored, however, now with an odd-parity GS. In terms of the charge parabolas this corresponds to odd-charge parabolas being always lower in energy than the even ones (red parabolas in Fig. 1h). This case of $2e$-periodicity for an odd-parity GS has not been reported before.

In the third case we illustrate the possible consequence of strong spin-orbit interaction. The zero-energy crossing of the subgap state can now be followed by a transition to a topological phase containing MZMs rigidly fixed at $E_0 = 0$ (Fig. 1e). (Note that the re-opening of the gap is not shown in Fig. 1e since only the lowest-energy subgap state is sketched.) As $E_0$ decreases below $E_c$, the peak spacing gradually evolves from $2e$ to $1e$ with alternating even-and odd-parity GSs (Figs. 1i and m). It is important to note that the even/odd degeneracy of the topological phase in bulk materials is lifted here by the charging energy [3,20]. This fundamental degeneracy is however visible by comparing the lowest energies for the even and odd parity GSs, which are both zero, albeit at different gate voltages. We note that ABSs confined by a smooth potential may also give rise to a similar phenomenon as sketched in Fig. 1e [21].

Finally, in the fourth case (Fig. 1f), the superconducting gap in Al closes at its critical perpendicular magnetic field (Figs. 1j and n). This transition to the normal state also causes equidistant 1$e$-periodic oscillations, which in the peak evolution is similar to the topological case. However, we show below that finite-bias spectroscopy is significantly different in these two cases.

In Fig. 2, we present exemplary data for the four cases illustrated in Figs. 1k-n. Fig. 2a-d show four panels of d$I$/d$V_b$ measured at zero bias as a function of $V_{PG}$ and $B$-field ($B_\parallel$ or $B_\perp$). The bottom row of panels shows representative linecuts at high $B$-fields. Fig. 2e-h show the corresponding peak spacings for even ($S_e$) and odd ($S_o$) GSs. These spacings are converted from gate voltage to energy via the gate lever arm and reflect the energy difference between even- and odd-parity states. In Fig. 2a the peaks are 2$e$-periodic with even-parity GS up to a field of ~0.9 T. This observation reflects that up to this $B$-field our Al thin film remains superconducting without any low-energy subgap state. Above ~0.9 T the gap is significantly suppressed such that 1$e$-transport sets in [11].

In Fig. 2b, the conductance peaks split into pairs around 0.11 T with alternating small and large spacings. These split peaks merge with neighbouring split peaks, leading to the recurrence of 2$e$-periodic oscillations, but strikingly with an odd-parity in the valleys (cf. Fig. 1l). The parity transition is also illustrated in Fig. 2f by the single crossing between $S_e$ and $S_o$. Similar to Fig. 2a, above 0.9 T the oscillation becomes 1$e$-periodic (see linecuts).

For the case shown in Fig. 2c (cf. Fig. 1m), the 2$e$-periodicity gradually changes to uniform 1$e$-periodicity above ~0.35 T. $S_e$ and $S_o$ exhibit slight but visible parity-changing oscillations up to 0.9 T, whose amplitude decreases with field. This case resembles the experiment of ref. [15] where the 1$e$ oscillations are associated with MZMs. Additionally, we found that the peaks are alternating in height. To quantify this effect, we extract from the data an asymmetry parameter $\Lambda = \frac{G_{e \to o}}{G_{e \to o} + G_{o \to e}}$, which amounts to 0.5 for peaks with equal heights [22]. Here, $G_{e \to o}$ ($G_{o \to e}$) is the peak height at an even-to-odd (odd-to-even) transition occurring upon increasing $V_{PG}$. Fig. 2g shows that $\Lambda$ undergoes drastic oscillations around 0.5 as $B_\parallel$ is varied, and levels off at 0.5 above 0.9T.

The data in Fig. 2d are taken for the same gate configuration as Fig. 2c but in a perpendicular $B$-field, which turns the Al into a normal state around $B_\perp$ = 0.18 T (see also Supplementary Fig. 3). In the normal state the oscillations are 1$e$-periodic and both $\Lambda$ and $S_e/S_o$ are constant, in agreement with established expectations [6].

The four columns in Figs. 1 and 2 represent distinct phases at high $B$: an even-parity GS, an odd-parity GS, a superconducting phase of alternating even and odd parities due to a single state at zero energy, and a gapless normal phase of alternating even and odd parities. These distinct phases can be reached by varying $V_{PG}$ and thereby the spatial profile of the wave functions, which determines the coupling strength to Al and to the external magnetic field [23-25]. In Fig. 2a, a very negative $V_{PG}$ pushes the wave functions against and partly into the Al, leading to a robust-induced gap with weak sensitivity to the $B$-field, which indeed never induces a parity change. Fig. 2b, at more positive $V_{PG}$, reflects the presence of a subgap state with larger weight in the InSb (as indicated by its estimated g-factor, $g \approx 7\text{-}15$, see another example in Fig. 3f). In Fig. 2c, at even more positive $V_{PG}$, the involved wave function has an even larger weight in the semiconductor, yielding a large $g$-factor of $\sim 10$, possibly augmented by orbital effects [26]. This leads to the appearance of robust zero-energy modes at $B$-fields much lower than the critical field of the thin Al shell. This $V_{PG}$-dependent 2e- to 1e-periodic transition has been repeated for another device, which is not depicted here.

**2$e$-periodic odd-parity GS**

Fig. 3 is dedicated to the new observation of a 2$e$-periodic, odd-parity GS. The combined system of a superconducting island weakly coupled to a quantum dot with an odd electron number can also have odd parity [27]. In our case all states are strongly hybridized with the superconductor and one bound state drops below $-E_c$, which causes the even-parity state to become an excited state above the odd-parity GS. (We describe different types of bound states in the Supplementary Part 5.) Fig. 3a-c show Coulomb diamond for the gate settings of Fig. 2b at three values of $B_\parallel$. Fig. 3a at $B_\parallel = 0$ shows 2$e$-periodic diamonds with 1$e$ and 2$e$ linecuts shown at the bottom (similar data was presented in Fig. 1b). Fig. 3b at $B_\parallel = 0.17$ T corresponds to the even-odd regime. Note that the conductance near $V_b = 0$ is suppressed, indicating that the subgap state causing the even-to-odd transition is weakly coupled to at least one of the normal leads (see also linecuts). Fig. 3c at $B_\parallel = 0.35$ T shows again 2$e$- periodic diamonds but now the GS inside the diamonds has an odd parity. The diamond structure, including the presence of regions with negative d$I$/d$V_b$, is very similar to the even-parity GS diamonds, except for the shift in gate charge by 1$e$.

Fig. 3f shows another example of the transition from the 2$e$-periodic even GS, via a region of even-odd spacings, to a 2$e$-periodic odd GS. Note again that the even-odd peak heights are significantly suppressed. These peaks correspond to a crossing of an even-parity parabola with an odd-parity parabola in Fig. 1h, where transport occurs via single electron tunnelling. In contrast, transport at the 2$e$-periodic peaks, both for even and odd GSs, occurs via Andreev reflection. The two cartoons (Fig. 3d and e) illustrate these different transport mechanisms.

**Isolated zero-energy modes and Coulomb valley oscillations**

Richer sequences of GS transitions as a function of magnetic field are also possible. For instance, the sketch in Fig. 4a illustrates the occurrence of multiple zero-energy crossings at low $B_\parallel$ followed by the appearance of a stable zero-energy state at higher $B_\parallel$. This type of behaviour is observed in Fig. 4b, showing large oscillations of $S_e$ and $S_o$ for $B_\parallel$ < 0.6 T, and a stable 1$e$-periodicity for $B_\parallel$ > 0.7 T (see Fig. 4c). Similarly, Fig. 4d also shows large oscillations of $S_e$ and $S_o$ below 0.6 T, followed by a region of almost equally-spaced peaks above 0.6 T (see Fig. 4e).

The features at low $B_\parallel$ (such as the position where the 2$e$-periodic peaks first split) can depend on the precise value of $V_{PG}$. In contrast, the features at high $B_\parallel$ are strikingly regular, with only a weak dependence on $V_{PG}$. This suggests that the 1$e$-periodicity at high fields originates from a state which is remarkably robust against gate variations. Furthermore, we note how the alternation of the conductance peak heights, already seen in Fig. 2c, is clearly visible in Fig. 4e even in the 1$e$-periodic regime. The origin of these peak height oscillations lies in the difference between tunnelling amplitudes involving the electron and hole components of the subgap states [22]. It was recently proposed that in an idealized model for MZMs in a finite-length wire, the oscillations of $\Lambda$ should be correlated with the oscillations in $S_e$ and $S_o$, i.e. that $\Lambda$ would be maximal or minimal when $S_e = S_o$ and vice versa that $|S_e - S_o|$ would be maximal for $\Lambda = 0.5$ [22]. In Fig. 4e, we find that the oscillations in $\Lambda$ are similar in number and period to the corresponding oscillations in $S_e$ and $S_o$, indeed suggesting a possible connection between the two (another example is presented in Supplementary Figs. 8c-f). Fig. 4f shows finite-bias spectroscopy in the 1$e$-periodic regime of Fig. 4d, at a high parallel field $B_\parallel = 0.7$ T. This spectroscopy reveals that the marked asymmetry of the peak heights originates from a discrete state that is gapped from a continuum of states at higher bias. As a comparison, Fig. 4g shows that for Al in the normal state no discrete features are observed. These are important verifications that substantiate our conclusion that the scenario in Fig. 1e is the proper description of the 1$e$ oscillations in the experimental figures (Figs. 2c and 4d) at high $B_\parallel$.

**Discussion**

In summary, we have revealed distinct types of fermion parity transitions occurring as a function of magnetic field and gate voltages in a Coulomb-blockaded InSb-Al island. These transitions provide a complete picture of all the parity phases in mesoscopic Cooper pair boxes [28]. Among these, in a finite field we find a novel odd-parity phase with 2$e$ periodicity in gate voltage. Additionally, we find 1$e$-periodic oscillations at high field originate from isolated zero-energy modes. The thorough understanding of the involved physics is important since such islands form the building blocks of future Majorana qubits [29-31].

**Methods**

**Device fabrication.**

An isolated Al segment is formed by selectively shadowing the nanowire during Al evaporation. InSb-Al nanowires with double shadows (Supplementary Fig. 1) were transferred from the InP growth chip to a doped-Si/SiO$_x$ substrate using a mechanical nanomanipulator installed inside an SEM. Au is used as leads and top gates. A 30 nm dielectric of SiN$_x$ separates the nanowire from the top gates.

**Transport Measurements.**

The device is cooled down to ~15 mK in an Oxford dry dilution refrigerator. The effective electron temperature is estimated to be 20-50 mK. Conductance across the devices was measured using a standard low-frequency lock-in technique (amplitude is 5 µV). The voltage bias $V_b = V_S - V_D$ is applied symmetrically between the two leads ($V_S = -V_D = V_b/2$). A magnetic field is applied using a 6-1-1 T vector magnet. The direction of the magnetic field is aligned carefully with respect to the nanowire axis (Supplementary Fig. 3). The sweeping rate of the datapoints is very slow and one $V_{PG}$-dependent linetrace at fixed $B$ in Fig. 2 takes a few minutes. We measure from 0 to 1T with 0.01T steps, bringing the measurement time for each panel to 3-4 hours. Because of this, we suffer from an ultra-slow drift that effectively changes the island potential. It is important to stress, however, that the slow drift affects the peak positions but not the peak spacings (since each trace only takes a few minutes to acquire). We extract subgap states and parity transitions from peak spacings and thus our conclusions are not affected by the ultra-slow drift.

**Data availability**

The figures are created from the row data. All data and the code used for peak fitting are available at http://doi.org/10.4121/uuid:e0ecafef-7f45-4475-b2f8-351c2af4a2b0 .

**Acknowledgements**

We gratefully acknowledge Joon Sue Lee, Daniel J. Pennachio and Borzoyeh Shojaei for help with superconductor/semiconductor fabrication and structural characterization, Petrus J. van Veldhoven for help with nanowire growth, and Kevin van Hoogdalem, Leonid Glazman and Roman Lutchyn for discussion of the data. This work has been supported by the European Research Council (ERC), the Dutch Organization for Scientific Research (NWO), the Office of Naval Research (ONR), the Laboratory for Physical Sciences and Microsoft Corporation Station Q.

**Author Contributions**

J.S. fabricated the devices. J.S., F.B. and S.J.J.R. performed the measurements. J.S., S.H., B.V.H. and L.P.K. analysed the data. G.W., D.X., D.B., and A.G. contributed to the discussion of data and the optimization of the fabrication recipe. S.G., R.L.M.O.H.V., D.C., J.A.L., M.P., C.J.P. and E.P.A.M.B. carried out the growth of materials. J.S., B.V.H., S.H., F.B. and L.P.K. co-wrote the paper. All authors commented on the manuscript.

**Competing interests**

The authors declare no competing interests.

**FIG. 1. Hybrid semiconducting-superconducting island and its energy spectrum**. **a**, False-colour scanning electron microscope image of the device consisting of an InSb nanowire (green) with an 800-900 nm long Al-shell (light-blue) covering the top facet and one side facet. Inset: schematic cross-section at the centre of the plunger gate (PG) indicated by the yellow line. The Si/SiO$_x$ substrate contains a global back gate that we keep at zero voltage. The InSb wire is contacted by Cr/Au leads (yellow) and then covered by a 30 nm thick dielectric layer of SiN$_x$ (light-grey). Ti/Au top gates (blue) that wrap around the wire allow for local electrostatic control of the electron density. Two gates are used to induce tunnel barriers (TG) and one plunger gate (PG) controls the electron number on the island. The scale bar indicates 500nm. **b**, d$I$/d$V_b$ versus tunnel gate voltage and $V_b$ showing 2$e$-periodic Coulomb diamonds (one diamond is outlined by yellow dashed lines). The lower panel shows horizontal linecuts with 2$e$-periodic Coulomb oscillations at $V_b$ = 0 (black trace) and 1$e$-periodic oscillations at $V_b$ = 150 μV (red trace). The panel on the right shows a vertical linecut through the Coulomb peak at the degeneracy point (blue trace) and through the centre of the Coulomb diamond (purple trace). Below are four scenarios for the $B$-dependence of a single Andreev level (**c**, **d**, **e** and **f**), the resulting energies as a function of the induced charge, $N_g$, (**g**, **h**, **i** and **j**) and the Coulomb oscillations (**k**, **l**, **m** and **n**). In panels **c-f**, the grey regions represent the continuum of states above Δ. The coloured traces represent the energy, $E_0$, of the lowest-energy subgap state. Panels **g-j** show the energies of the island with $N$ excess electrons, $E(N_g) = E_c (N_g - N)^2 + p_N E_0$, where $N_g$ is the gate-induced charge, $N$ is the electron occupancy number, and $p_N$ = 0 (1) for $N$ = even (odd). Parabolas for $N$ = even are shown in black, while parabolas for $N$ = odd are shown in colours in correspondence to the colours in the other rows. Crossings in the lowest-energy parabolas correspond to Coulomb peaks as sketched in panels **k-n**, again with the same colour coding. Labels in the Coulomb valleys between the peaks indicate the GS parity being either even (e) or odd (o).

**FIG. 2. Four representative evolutions of Coulomb peaks**, corresponding to the four columns (**c-n**) in Fig. 1. Top row panels: d$I$/d$V_b$ as a function of $V_{PG}$ and $B_\parallel$ (**a**, **b**, **c**) or $B_\perp$ (**d**). Below are typical linecuts at different $B$-fields indicated by the purple and green lines. (**e**, **f**, **g**, and **h**) Even and odd peak spacings, $S_e$ (red) and $S_o$ (blue) on the left axis, and peak height ratio, $\Lambda$ (black) on the right axis, versus $B$-field, for the valleys labelled e/o in **a** and **b** and for the average spacings in **c** and **d**, respectively. Here and in other figures, the linewidths of $S_e$ and $S_o$ curves correspond to 5 µeV, in accordance with the lock-in excitation energy. **a**, The 2$e$-periodicity with even-parity valleys persists up to 0.9 T, above which quasiparticle poisoning occurs. **b**, The 2$e$-periodic peaks split at ∼0.11 T and merge again at ∼0.23 T. For $B_\parallel$ > 0.23 T, the oscillations are again 2$e$-periodic, but here the GS parity is odd, consistent with Fig. 1h. **c**, 2$e$-periodicity transitioning to uniform 1$e$-periodicity at $B_\parallel \approx 0.35$ T, accompanied by peak spacing and peak height ratio oscillations up to 0.9 T (see also panel **g**). **d**, 2$e$-periodicity transitioning to 1$e$-periodicity at $B_\perp$ = 0.18 T (the vertical dashed line), coinciding with the critical $B$-field of the Al layer (see Supplementary Fig. 3). Above the critical field, peak heights are constant (see linecuts) and the even/odd peak spacings are equal (**h**). A few common offsets in $V_{PG}$ are introduced to compensate the shifts in gate voltage, and the raw data are listed in Supplementary Fig. 4.

**FIG. 3. Transport via an odd-parity GS.** (**a-c**) Coulomb diamonds for the gate settings in Fig. 2b at different $B_\parallel$. Below are linecuts at $V_b$ = 0 (black trace) and $V_b$ = 150 µV (red trace),

respectively. **f**, Another example of an even- to odd- GS transition. The inset is a zoom-in of the even-odd regime with a different scale bar. The sketch in **d** illustrates the Cooper pair tunnelling process in both even- and odd- GS regimes (marked by green dashed lines in **f**), while **e** illustrates the single electron tunnelling process in the alternating even-odd parity GS regime (marked by yellow dashed lines in **f**).

**FIG. 4. Evolution of multiple subgap states. a**, Schematic $B$-field dependence for the case of three subgap states with GS parity transitions at each zero-energy crossing. Dashed lines indicate level repulsion between different subgap states, leading to large oscillations of the lowest energy $E_0$. **b**, One example of Coulomb peaks reflected by the scenario in panel **a**. The extracted peak spacings for the valleys labelled by e and o are shown below in **c**. The evolution of the peak spacings is compatible with the type of energy spectrum shown in **a**, characterized by large oscillations of $E_0$. Note that the odd-parity GS around ~0.2 T develops a full 2$e$-periodicity. (The spacings near ~0.6 T are absent because the exact peak positions are unclear.) **d**, Another example of Coulomb oscillations with a pronounced $B$-field dependence. Extracted peak spacings and the height ratio are shown in **e** (averaged over the three periods in **d**). **f**, Coulomb diamonds at $B_\parallel$ = 0.7 T along the $V_{PG}$-range indicated by the yellow line in **d.** This bias spectroscopy reveals isolated zero-bias peaks at the charge-degeneracy points that are separated from the continuum. As in Fig. 2c, neighbouring zero-bias peaks have different heights (also visible in **d** at high $B_\parallel$). **g**, Bias spectroscopy with Al in the normal state ($B_\perp$ = 0.3 T) where the isolated zero-bias peaks are absent.

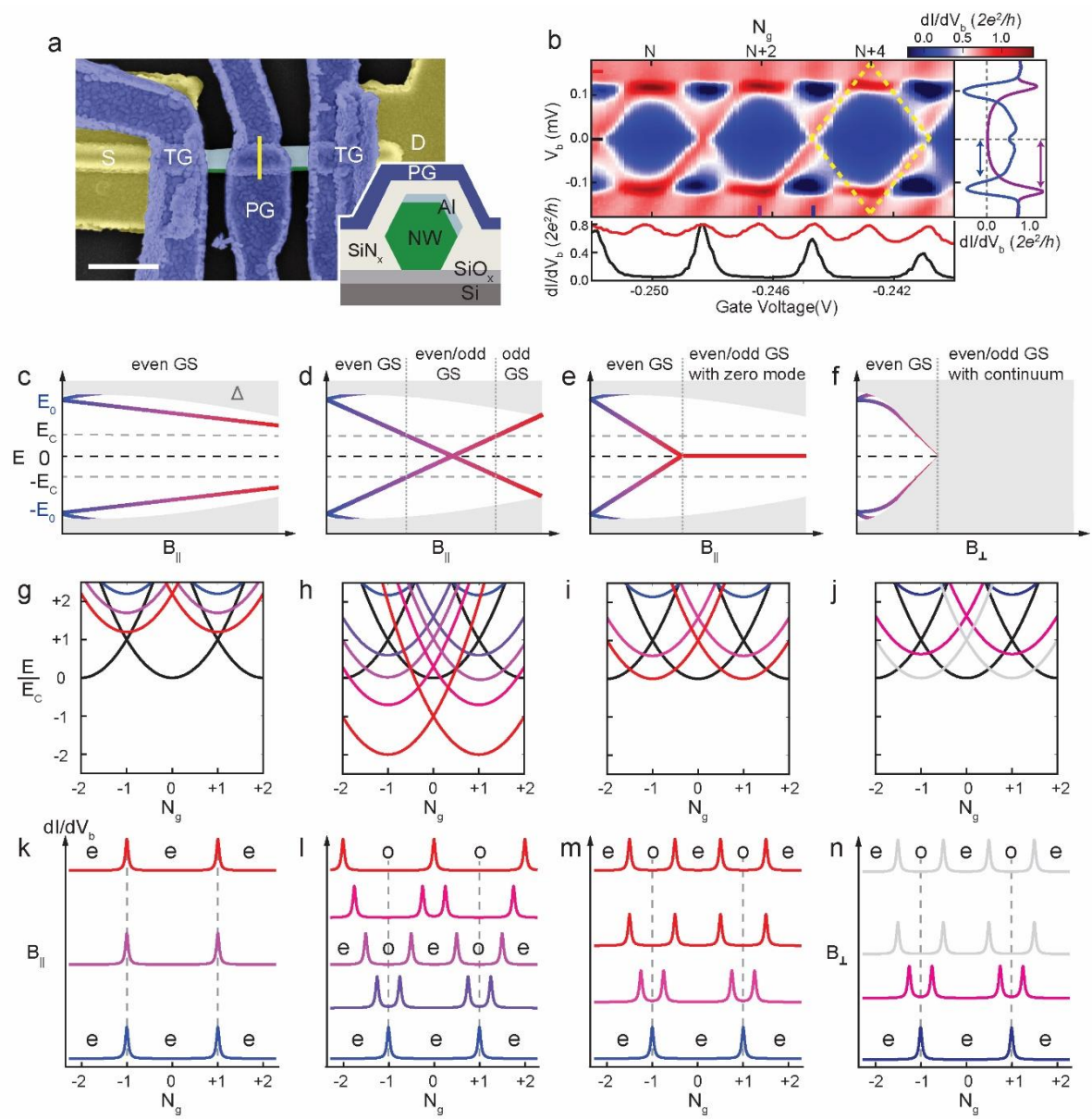

**Figure 1**

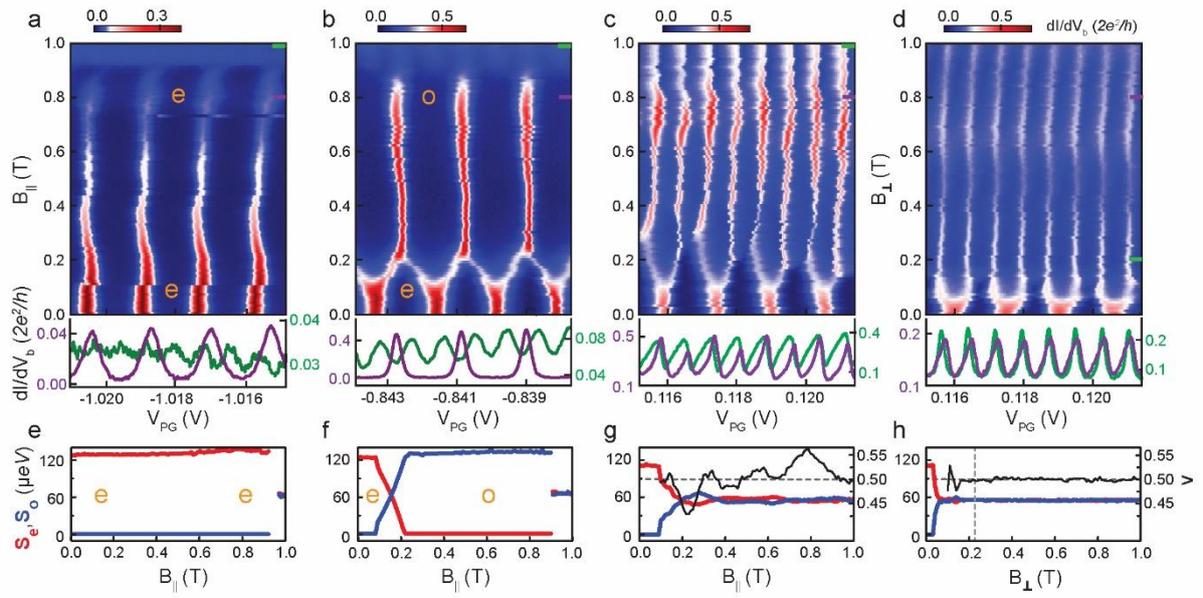

**Figure 2**

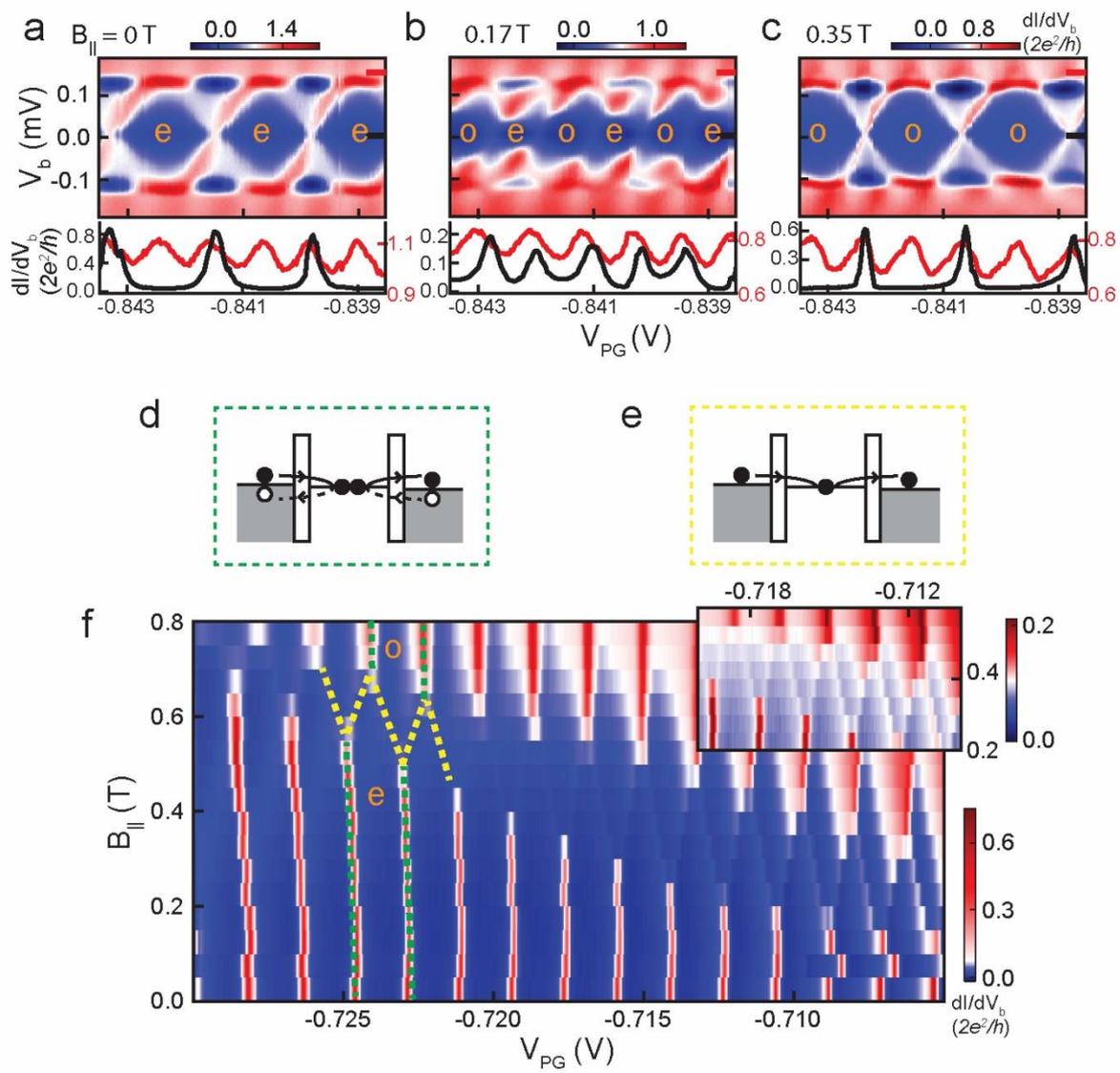

**Figure 3**

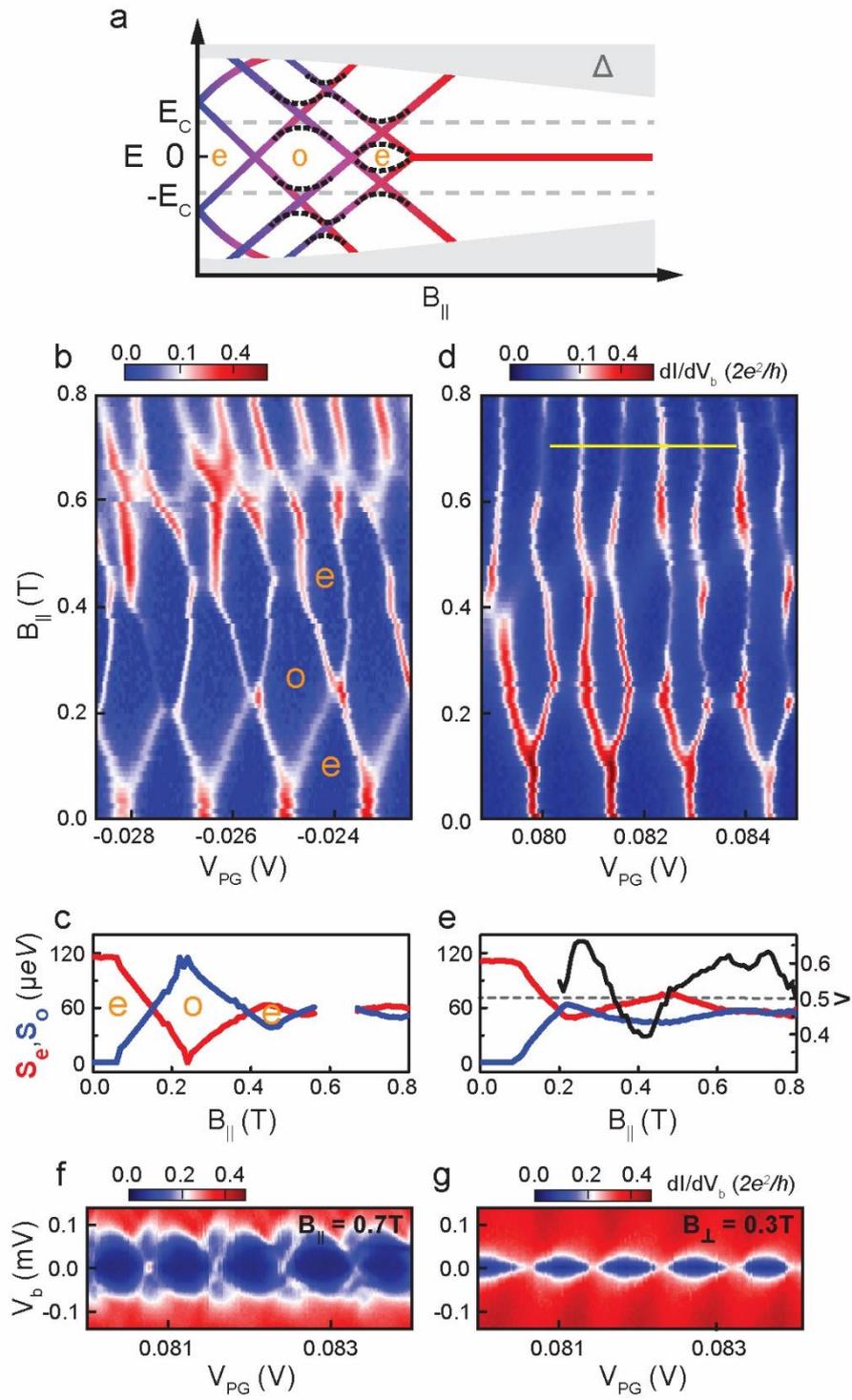

**Figure 4**

**Supplementary Note 1. Growth of epitaxial InSb/Al nanowire islands**

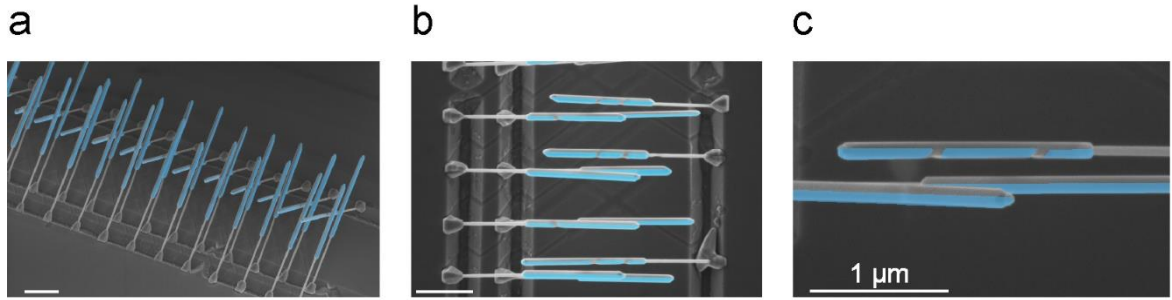

**Supplementary Figure 1. Scanning electron microscope (SEM) image of shadowed InSb/Al nanowires.** The Al shell on the nanowires is coloured in light blue. **a**, InSb nanowires are grown in a metal-organic vapour-phase epitaxy (MOVPE) reactor from catalyst droplets (Au), positioned along etched trenches. First, InP stems are grown to facilitate the nucleation of InSb nanowires. After nanowire growth, hydrogen cleaning is used to remove the native oxide layer of the nanowires and Al is evaporated in a direction parallel to the trenches. The positioning of Au droplets together with the tilting of the Al evaporation direction with respect to the horizontal plane allows the shadowing of predefined sections on the nanowires. **b**, Zoomed-in SEM image of InSb nanowires with either one or two shadows. **c**, During the Al evaporation, two sacrificial nanowires at the bottom are used to shadow two short segments (∼100 nm) on a third wire at the top. The island length, set by the separation between two shadows, varies between 0.2 and 1 μm. Note that this 'shadow-growth' mechanism avoids the need of etching the Al, leaving pristine nanowire facets on the junction regions. We create a hybrid superconducting-semiconductor island of Al-InSb by interrupting the Al shell in two narrow regions allowing local electrostatic gating of two semiconducting junctions. At these two regions tunnel barriers can be introduced by the top gates to confine the superconducting-semiconducting hybrid island.

**Supplementary Note 2. Extracting $E_c$ and $E_0$ from 2$e$-periodic Coulomb diamonds at zero magnetic field**

The energy parabolas in Supplementary Figure.2a illustrate that the degeneracies of the even-parity ground-state parabolas occur at $E_c$. In the Coulomb valley (at $N_g = 0$) the even-parity parabolas cross at $4E_c$ (indicated by the yellow dot in Supplementary Figure.2a). In finite-bias Coulomb diamonds, the voltage drop at the top of the 2$e$-periodic Coulomb diamonds corresponds to $8E_c/e$ (dashed yellow diamond in Supplementary Figure.2b). As a result, $E_c$ extracted from the yellow-dashed diamond in Fig. 1b of the main text is ∼22 μeV. Over the entire gate range $E_c$ varies between 22-27 μeV (see typical diamonds in Supplementary Figure.2c and d at different gate values).

Above the degeneracy points of the GSs, the onset of quasiparticle transport causes a blockade of Andreev reflection and results in a region of negative differential conductance (NDC) starting from a threshold voltage bias $V_{NDC} \approx 2(E_0 - E_c)/e$ (see the blue arrows in Supplementary Figure.2a) [2]. For example, $V_{NDC} = 90$ μV in Fig. 1b, so $E_0 \approx 67$ μeV. On the other hand, at finite bias, the onset of 1$e$-periodicity is due to single-particle transport via co-tunnelling events (see the red arrow in Supplementary Figure.2a), corresponding to $2E_0/e$ [3]. In Fig. 1b, $E_0$ extracted from the onset of 1$e$-periodicity is close to the number extracted from NDC in the same diamond ($E_0 \approx 67$ μeV). Over the entire gate range $E_0$ varies between 50 - 90 μeV (see diamonds in Supplementary Figure.2c and d).

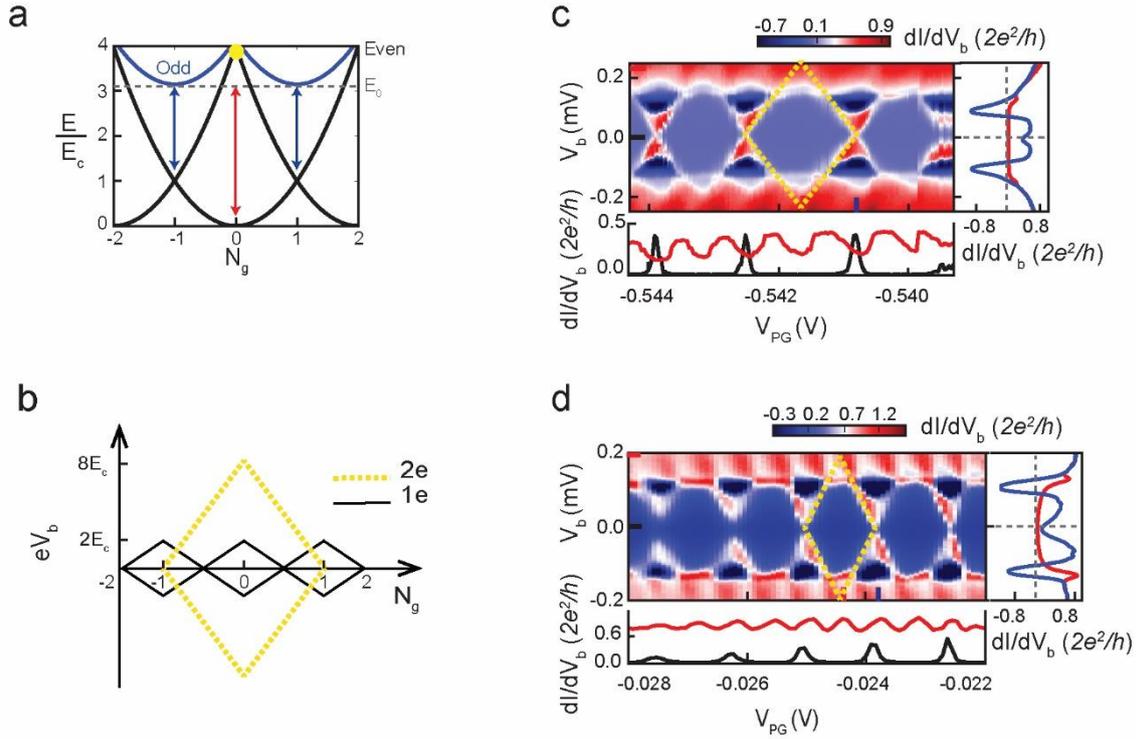

**Supplementary Figure 2. Derivation of $E_c$ and $E_0$ from Coulomb diamonds. a**, The energy level parabolas of the superconducting island. The black parabolas describe the charge states for even parity, and blue parabolas for odd parity. Odd parabolas are lifted by $E_0$, consistent with $E_0 \gg E_c$ in Fig. 1b. **b**, Finite-bias Coulomb diamonds for 1$e$ (solid black diamonds) and 2$e$-periodicity (dashed yellow diamond). **c** and **d**, Coulomb diamonds for two different gate configurations. In the bottom panels, horizontal linecuts show the 2$e$ (in black) versus 1$e$-periodic (in red) conductance oscillations taken respectively at $V_b$ = 0 µV and $V_b$ = 250 µV (in **c**) / $V_b$ = 200 µV (in **d**). In the right panels, vertical linecuts at different $V_{PG}$ voltages show the presence of NDC regions above the degeneracy point (blue linecuts) and conductance enhancement in the valley (red linecuts). For **c**, we estimate $E_c \approx 25$ µeV, while the onset of NDC is found at $V_b$(NDC) ≈ 70 µV, so $E_0 \approx 60$ µeV. In **d**, $E_c \approx 22$ µeV and $V_b$(NDC) ≈ 70 µV, so $E_0 \approx 57$ µeV.

**Supplementary Note 3. Superconducting critical magnetic fields of the InSb/Al nanowire for three orientations**

We performed tunnelling spectroscopy at one of the two junctions as a function of the magnetic field strength in three directions: parallel to the nanowire $B_{\parallel}$ (Supplementary Figure.3a), perpendicular to the substrate $B_{\text{out of plane}}$ (Supplementary Figure.3b) and perpendicular to the nanowire in the plane of the substrate $B_{\text{in plane}}$ (Supplementary Figure.3c). The local tunnel gate voltage is -1 V for this spectroscopy junction (the weak-tunnelling regime). The voltages are +2 V for the other tunnel gate and the plunger gate to make sure the chemical potential is smooth for the entire island except at the local tunnel gate. When $B$ is applied parallel to the nanowire (Supplementary Figure.3a), the hard superconducting gap persists up to 1.0 T, with Δ($B_{\parallel}$ = 0) = 220 µeV and Δ($B_{\parallel}$ = 0.8 T) = 90 µeV (the bottom panel in Supplementary Figure.3a). The gap closes completely at $B_{\parallel}^c \geq 1$ T, which is out of range for the employed 3D vector magnet.

For the other two orientations, our device undergoes a transition to the normal state at $B_{\text{out of plane}}^c \approx 0.12$ T and $B_{\text{in plane}}^c \approx 0.18$ T. In the main text, $B_{\perp}$ means $B_{\text{in plane}}$ for simplicity.

This observation is consistent with the SEM images of the device in Fig. 1a, showing the Al shell covering the top facet and one of the side facets of the nanowire.

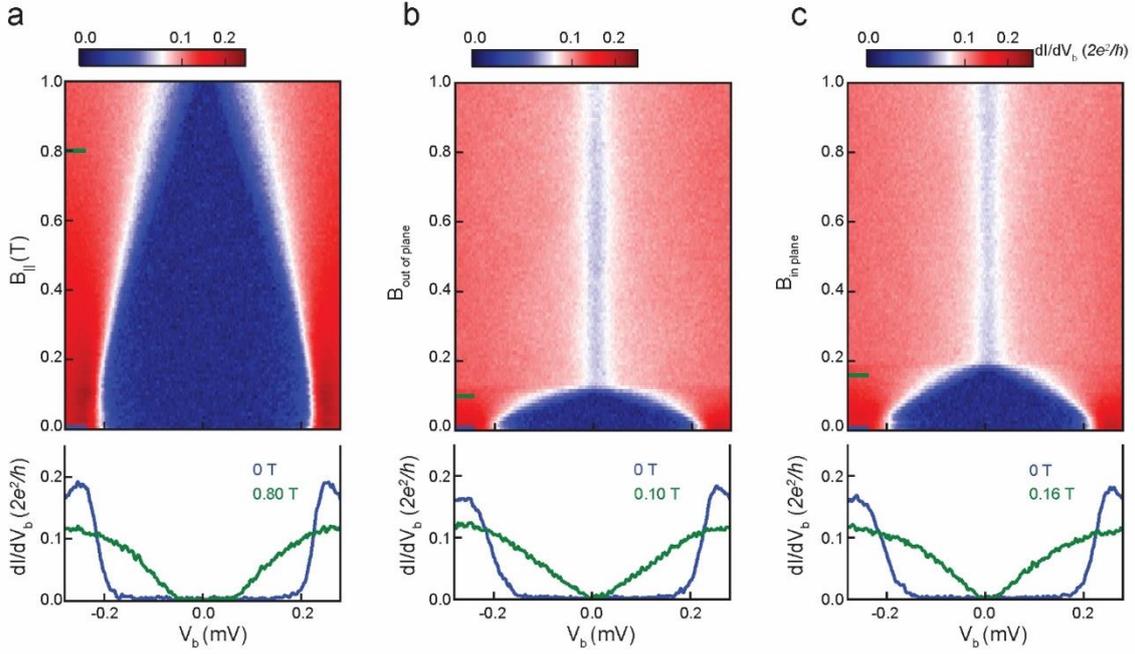

**Supplementary Figure 3. Tunnelling spectroscopy at magnetic fields of different orientations.** Top panels show d$I$/d$V_b$ tunnelling spectroscopy as a function of $B_\parallel$ (**a**), $B_{\text{out of plane}}$ (**b**) and $B_{\text{in plane}}$ (**c**). Typical linecuts at selected $B$ are shown in the bottom panels. The zero-bias dip above the critical magnet field in **b** and **c** are likely due to the confinement at the second junction.

## Supplementary Note 4. Raw data for d$I$/d$V_b$ as a function of $V_{PG}$ and $B_\parallel$

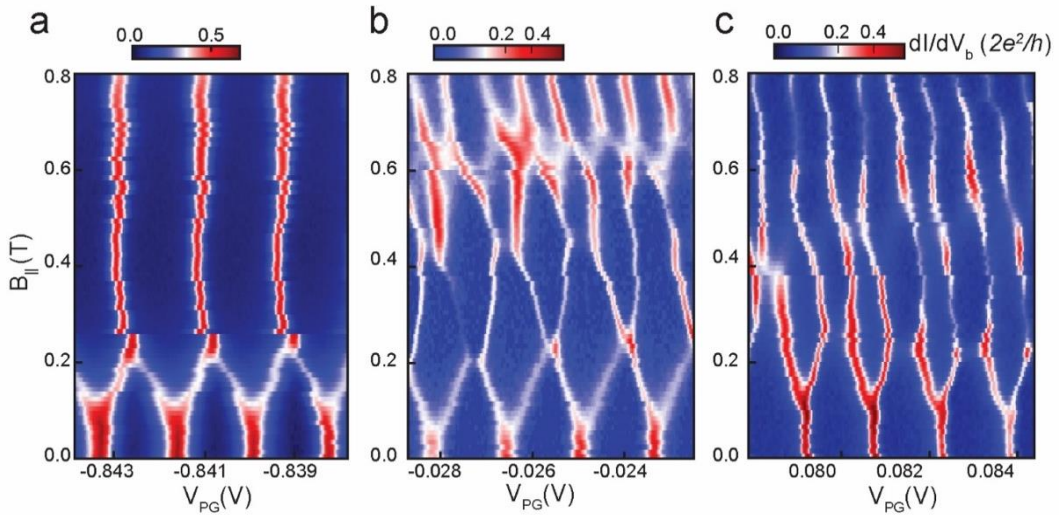

**Supplementary Figure 4. Raw data for d$I$/d$V_b$ as a function of $V_{PG}$ and $B_\parallel$. a**, **b**, and **c** correspond to Fig. 2b, Fig. 4b and Fig. 4d in the main text, respectively. The plunger gate is sometimes drifting and gate-voltage jumps can occur because of charge trapping in the dielectric, so that raw d$I$/d$V_b$ data here are not always stable. However, the conductance peaks from different parity states are still easily identified. A few common offsets in VPG are introduced to compensate for the shifts in gate voltage.

**Supplementary Note 5. Comparison of the 2e-periodic odd-parity ground state with previous observations of fermion parity crossings**

The ground state fermion parity of a superconducting system can be changed by the Fermi-level crossing of a spin-resolved subgap state. The subgap state causing the GS transition may, for instance, be bound to an impurity in a bulk superconductor, as in the case of Shiba-Yu-Rusinov states [4]; or it may be an Andreev bound state in a Josephson junction (see [5] for instance). In our case, the fermion parity switch is due to a subgap state localized in the mesoscopic InSb/Al superconducting island. In this case, as discussed in the main text, when the energy gap for emptying the subgap state exceeds the charging energy of the island, the odd parity GS caused by the fermion parity switch is stable at all value of the induced charge of the island and the sequential process is Cooper pair tunnelling/Andreev reflection. This condition was not reached in previous studies of hybrid superconducting-semiconductor islands, which always observed either an even-parity GS or an alternation of even- and odd-parity GS [6, 7]. Furthermore, our observation is also distinct from the parity transitions observed in semiconducting quantum dots proximitized by a superconducting lead [8], where the charging energy applies only to the semiconductor but not to the superconductor. In this case, reducing the strength of the proximity effect (by varying the coupling between the dot and the superconductor) may cause a change from even to odd parity in the ground state of the dot, but only at values of the induced charge which would favour an odd occupation of the dot in the absence of the superconducting lead. In other words, the effect does not require a change of the GS parity of the superconducting lead itself.

**Supplementary Note 6. Effect of parallel and perpendicular magnetic fields on the periodicity of Coulomb peaks**

For Figs. 2c and 4d of the main text, we just show typical Coulomb diamonds (Fig. 4f) at a specific value of $B_\parallel$. The additional figures at different $B_\parallel$ and $B_\perp$ (presented in Supplementary Figure.6) demonstrate the robustness of the isolated zero-bias state at all $B_\parallel$, as well as the obvious difference between the superconducting state at $B_\parallel$ and the normal state at $B_\perp$. Supplementary Figures.6a-c correspond to the same gate settings as Fig. 4d and Supplementary Figure.6d relates to Figs. 2c-d. The diamonds at different $B_\parallel$ (Supplementary Figure.6b and the top two panels in Supplementary Figure.6d) show the consistence of the isolated zero-mode. The finite bias spectroscopy (Supplementary Figure.6a) at the degeneracy point of Fig. 4d, as well as Supplementary Figure.6b, proves there is a zero-energy crossing at low $B_\parallel$ and a sticking zero-energy state at high $B_\parallel$, which fits the sketch in Fig. 4a. The normal transition in Supplementary Figure.6c shows equal peak spacings and heights, which is used to distinguish the normal and superconducting 1*e*-periodic Coulomb peaks. For figs. 2c and d, we can also see the isolated zero-mode for different $B_\parallel$ and a continuum in the normal regime (Supplementary Figure.6d).

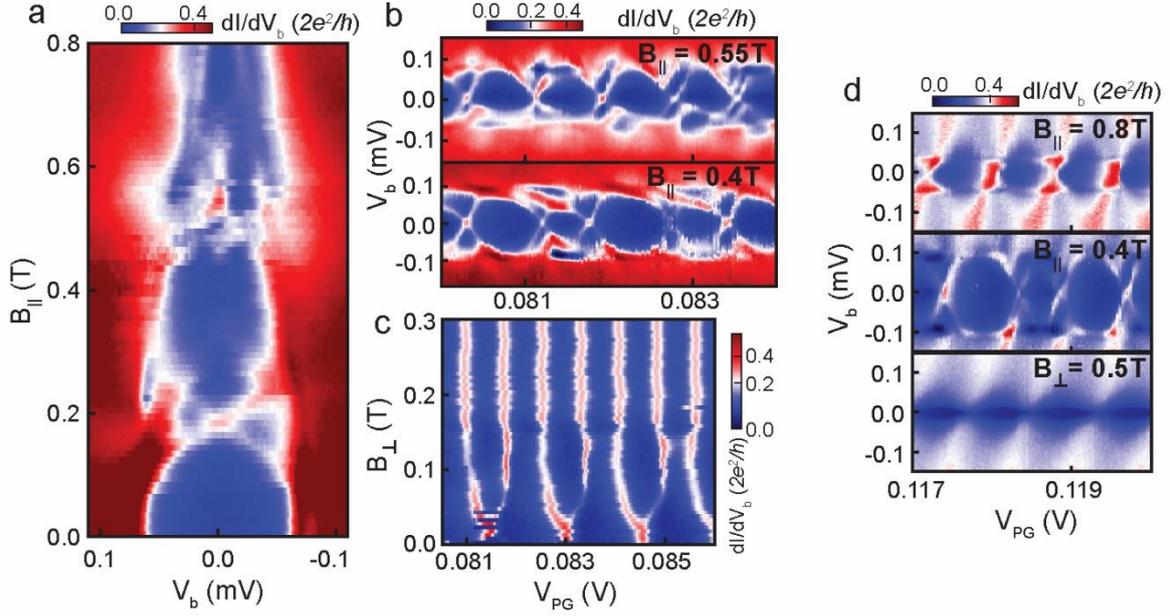

**Supplementary Figure 6. Zero-bias resonances at different $B_\parallel$ and $B_\perp$. a, b** and **c** are measured for a similar gate-voltage regime as Fig. 4d. **a**, $dI/dV_b$ as a function of $V_b$ and $B_\parallel$ at $V_{PG}$ close to one degeneracy point in Fig. 4d. A charge degeneracy point at first crosses the Fermi level at $B_\parallel \approx 0.2$ T, while a more persistent zero-bias peak occurs for $B_\parallel$ = 0.5 - 0.65 T. **b**, Coulomb diamonds at $B_\parallel \approx 0.4$ T and 0.55 T. Both of them show a discrete state at the degeneracy points, isolated by an energy gap. $G_{e \to o}$ and $G_{o \to e}$ also show alternating amplitudes. **c**, Evolution of zero-bias conductance peaks with $B_\perp$. The state becomes normal at $B_\perp \approx 0.18$T ($B^c_{\text{in plane}}$ in Fig. S3c), and the peak oscillations become 1e-periodic with equal peak heights. **d**, Coulomb diamonds at different $B_\parallel$ and $B_\perp$ at the same gate-voltage regime as Figs. 2c and 2d. Both of the top and middle panels at finite $B_\parallel$ show a discrete level at the degeneracy points and alternating peak heights, whereas the diamonds in the normal regime in the bottom panel show normal 1e oscillations without isolated peaks at the charge degeneracy points.

## Supplementary Note 7. Fitting of the Coulomb resonances

The Coulomb resonances are analysed by fitting all peaks simultaneously using an identical electron temperature $T_{el}$ that takes into account the temperature-broadening of the Coulomb resonances. A single resonance is described by a Breit-Wigner distribution [9, 10]

$$G_{BW}(V_{PG}, V_0, E) = \frac{2e^2}{h} \frac{(h\Gamma/2)^2}{(h\Gamma/2)^2 + [e\alpha(V_{PG} - V_0) - E]^2} \quad (1)$$

where $V_0$ is the centre of the Coulomb peak and $\alpha$ is the plunger gate lever arm. The line shape is thermally broadened according to [11]

$$G(V_{PG}, V_0) \propto \int_{-\infty}^{\infty} G_{BW}(V_{PG}, V_0, E) \left[ -\frac{\partial f(T, E)}{\partial E} \right] dE \quad (2)$$

with the Fermi-Dirac distribution $f(T, E)$. Hence, the total fitting function is given by a sum over a number of these line shapes and the tunnel coupling to the leads $\Gamma$. The peak position $V_0$ and the peak height are individual fitting parameters for each resonance, while the electron temperature $T_{el}$ and a constant offset are used as common fitting parameters.

In conclusion, by fitting the data as depicted exemplarily in Supplementary Figure.7a (blue) the fitted curve (green) describes the data very well and we find an electron temperature of about 20-50 mK and a typical tunnel coupling of about $h\Gamma = 5$ µeV.

As depicted in Fig. 4b (bottom panel) of the main text the Coulomb peak spacing of the even and odd valleys oscillates as a function of magnetic field. The oscillation amplitude is strongly reduced above 0.6 T. In Fig. 4b also the relative Coulomb peak height $\Lambda$ is shown averaged for the three pairs of Coulomb peaks in Supplementary Figure.6b. This quantity is given by

$$\Lambda = \frac{G_{e \to o}}{G_{e \to o} + G_{o \to e}} \quad (3)$$

where $G_{e \to o}$ is the conductance peak height at resonance between an even and an odd parity Coulomb diamond and $G_{o \to e}$ is the consecutive resonance between the odd and the next even parity Coulomb diamond [12]. Clearly, the relative peak height undergoes oscillations as well that extend also into the regime of stable 1e oscillations above 0.6 T. The fitting result for the data in Supplementary Figure.7b, which is used to extract the Coulomb peak spacings and peak heights, is depicted in Supplementary Figure.7c.

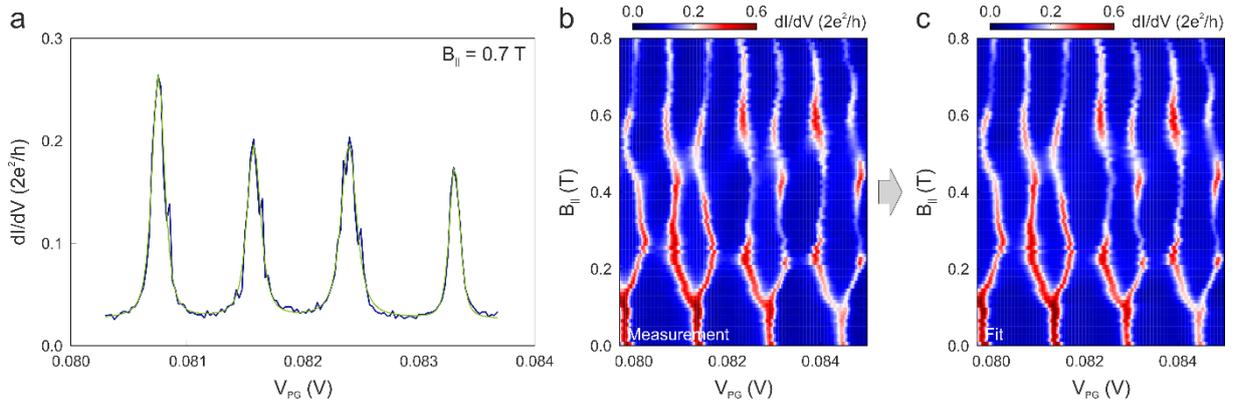

**Supplementary Figure 7. Fitting of the Coulomb resonances. a**, Fitting of the zero-bias linecut at $B_\parallel = 0.7$ T from the data presented in Fig. 4d of the main text. **b** and **c**, Measurement (**b**) and fitting result (**c**) using $G_{\text{sum}} = \sum_i G(V_{\text{PG}}, V_{0,i})$, with $G(V_{\text{PG}}, V_{0,i})$ given by Supplementary Eq.(2), for the data presented in Fig. 4d of the main text.

## Supplementary Note 8. The relation between peak spacing and peak height ratio

We observe that the oscillations in $\Lambda$ are similar in number and period to the corresponding oscillations in $S_e$ and $S_o$ (Supplementary Figure.8a, b and d), indeed suggesting a possible connection between the two.

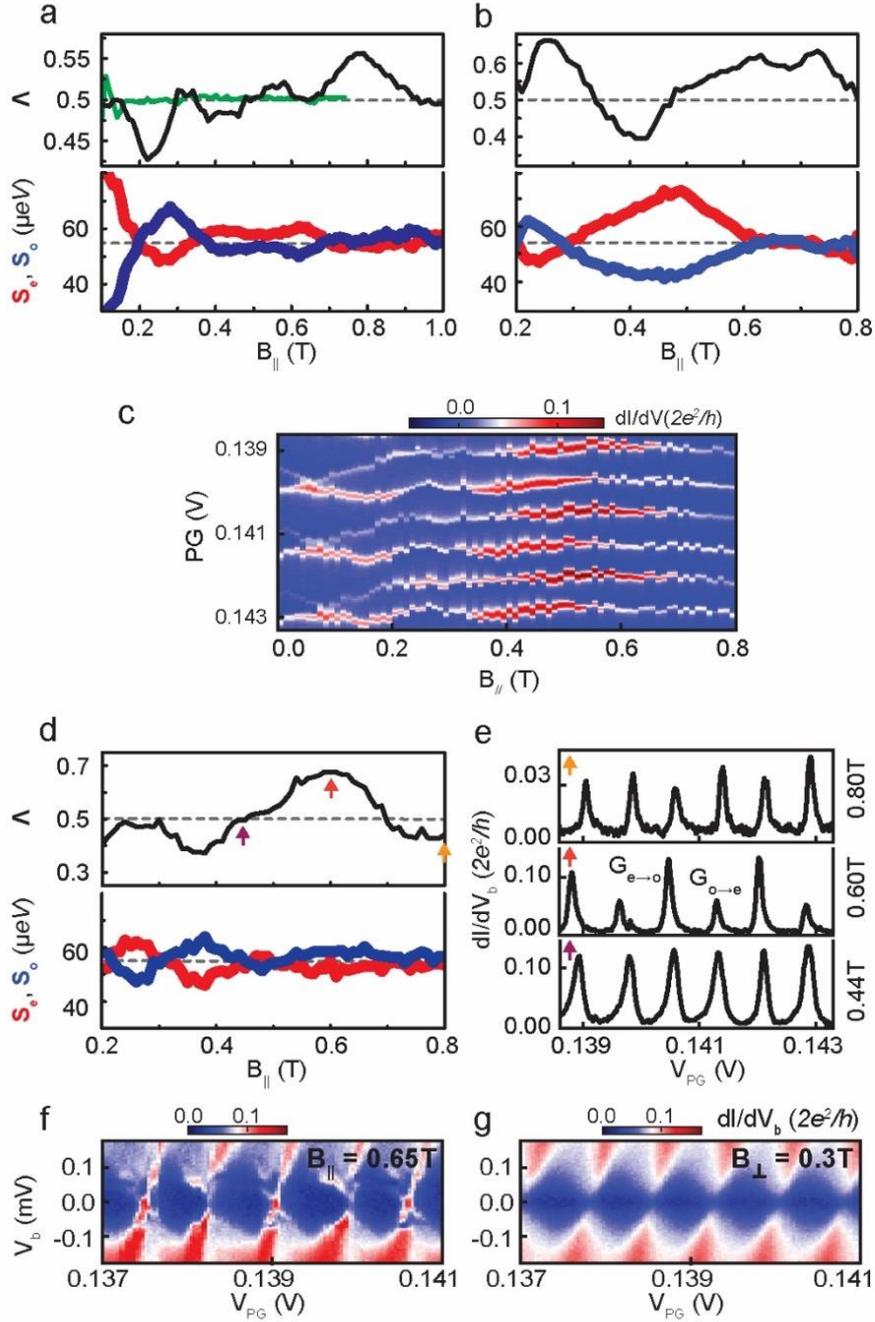

**Supplementary Figure 8. Evolution of peak heights and spacings.** Peak height ratio $\Lambda$ and peak pacings for even ($S_e$) and odd ($S_o$) parities as a function of $B_\parallel$ extracted from Fig. 2c (shown in **a**), Fig. 4d (shown in **b**) and Fig. S8c (shown in **d**). For comparison, the green curve in **a** is extracted from the normal state data of Fig. 2d. **e**, Exemplary Coulomb oscillations for the data in **c** at different fields $B_\parallel$ = 0.44 T ($\Lambda \approx 0.5$), 0.60 T ($\Lambda > 0.5$), and 0.80 T ($\Lambda < 0.5$) indicated by arrows. **f**, Coulomb diamonds at a large value of $\Lambda$ in **c**, indicating the isolated zero mode. **g**, Coulomb diamonds in the same gate-voltage regime as **f**, but measured in the normal state ($B_\perp$ = 0.3 T), showing the continuum at finite bias.

## Supplementary Information References